\documentstyle[prl,aps,multicol,epsf]{revtex}
\begin{document}
\draft
\title{Mott Transition in Degenerate Hubbard Models:  
       Application to Doped Fullerenes}

\author{Olle Gunnarsson$^{(a)}$, Erik Koch$^{(a),(b)}$,
        and Richard M. Martin$^{(a),(b)}$}
\address{${}^{(a)}$Max-Planck-Institut f\"ur Festk\"orperforschung,
         D-70506 Stuttgart, Germany}
\address{${}^{(b)}$Department of Physics, University of Illinois at
         Urbana-Champaign, Urbana, Illinois 61801}
\date{April 22, 1996}
\maketitle

\begin{abstract}
The Mott-Hubbard transition is studied for a Hubbard model with orbital 
degeneracy $N$, using a diffusion Monte Carlo method. Based on general 
arguments, we conjecture that the Mott-Hubbard transition takes place for 
$U/W\sim \sqrt{N}$, where $U$ is the Coulomb interaction and $W$ is the 
band width. 
This is supported by exact diagonalization and Monte Carlo calculations. 
Realistic parameters for the doped fullerenes lead to the conclusions that
stoichiometric A$_3$C$_{60}$ (A=K, Rb) are near the Mott-Hubbard
transition, in a correlated metallic state.           
\end{abstract}
\pacs{71.10.Fd, 71.30.+h, 71.20.Tx\hfill cond-mat/9608107}

\vspace{-3ex}
\begin{multicols}{2}
\narrowtext
The fullerenes have attracted much interest, not least because of the 
superconductivity of A$_3$C$_{60}$ (A=K, Rb)\cite{sc}. Much work has been 
performed to explain this phenomenon\cite{Varma,Schluter,Mazin,Antropov,mu,C60-}
starting from the assumption that the systems are metallic. The metalicity 
is an experimental fact\cite{metal}, but surprising from a theoretical point 
of view. The Coulomb interaction $U$ between two electrons on the same C$_{60}$
molecule may be a factor 1.5-2.5 larger than the width $W$ of the (partly 
occupied) $t_{1u}$ band\cite{Sawatzky,Bruhwiler,UAntropov,Pickett,Martins,%
WAntropov,Louie}. Since systems with $U/W$ exceeding a critical value a bit 
larger than unity ($\sim 1.5$) are believed to be Mott-Hubbard insulators\cite{%
Mott-Hubbard}, one would expect this to apply to doped C$_{60}$ compounds.
It has therefore been suggested that stoichiometric K$_3$C$_{60}$ is a Mott 
insulator with a band gap of 0.7 eV, and that experimental samples are metallic
only because they are nonstoichiometric\cite{Sawatzky}. Alternatively, the 
critical value $U_c/W$ could be larger than commonly believed. 
Which one of these alternatives apply, could have important implications for 
the properties of nonstoichiometric A$_3$C$_{60}$ as well.
We here study the Hubbard model, which has become the standard model for 
describing correlated systems. In spite of the extensive studies of this model,
$U_c/W$ is known accurately only for infinite dimension in the absence of 
orbital degeneracy\cite{Mott-Hubbard}. Since, however, almost all systems of 
physical interest have orbital degeneracy\cite{Bickers}, we here focus on the 
effects of such a degeneracy, considering a three-dimensional fcc lattice.

First, we provide qualitative arguments suggesting that for a half-filled 
system with orbital degeneracy $N$, the critical value of $U/W$ is enhanced 
by a factor $\sqrt{N}$. These arguments are supported by exact diagonalization 
calculations for small systems and  by  diffusion Monte Carlo calculations for 
larger systems. The reason for the  enhancement is that in the many-body 
treatment, the degeneracy increases the hopping rate of an extra electron or 
hole in these systems relative to the one-particle band width. The degeneracy 
can thus stabilize a metallic state which may be highly correlated.
This puts the doped fullerenes A$_n$C$_{60}$ ($n=$3, 4) ($N=3$) close to a
Mott-Hubbard transition, so that a correlated metallic state of stoichiometric
A$_3$C$_{60}$ is consistent with the large values of $U/W$ inferred from from 
experimental and theoretical information.

In a treatment using the Gutzwiller Ansatz and the Gutzwiller 
approximation\cite{Gutzwiller} for the Hubbard model, Lu found the critical 
value $U/W$ to vary as $(N+1)$ at half-filling\cite{Lu}. Here we provide 
qualitative arguments showing that the degeneracy is important, but that the 
critical value is substantially smaller and has a weaker ($\sqrt{N}$) 
dependence than in the work of Lu. Our arguments are supported by exact 
diagonalization and Quantum Monte-Carlo calculations.

To discuss the Mott-Hubbard gap for a model with the orbital degeneracy $N$, 
we first consider a (bipartite) lattice with $M$ sites and a hopping integral 
$t_{im,jm'} =t\delta_{mm'}$ between orbitals with the same quantum numbers $m$ 
on neighboring  sites $i$ and $j$. For half-filling the energy gap is then 
\begin{equation}\label{eq:1}
E_g=E(NM+1)+E(NM-1)-2E(NM),
\end{equation}
where $E(L)$ is the ground-state energy for $L$ electrons. For $U\gg t$ we have
\begin{equation}\label{eq:2}
E(NM)={1\over 2}N(N-1)MU +O({t^2\over U}),
\end{equation}
since no electron can hop without an extra energy cost $U$. For instance, this 
energy is obtained for a Neel state $|0\rangle$, with the moment $N$. In the 
system with $NM+1$ electrons, the extra electron can hop without an extra 
energy cost $U$. We thus consider the states
\begin{eqnarray}\label{eq:3}
&&|1\rangle=\psi^{\dagger}_{11\downarrow}|0\rangle      \nonumber  \\
&&|i\rangle={1\over \sqrt{N}}\psi^{\dagger}_{11\downarrow}
             \sum_m\psi^{\dagger}_{im\uparrow} \psi_{1m\uparrow}|0\rangle  ,
\end{eqnarray}
where $\psi^{\dagger}_{jm\uparrow}$ creates a spin up electron on site $j$ in 
orbital $m$. The sites $i$ are nearest neighbors of site 1, which we have 
assumed to have spin down electrons in the state $|0\rangle$. We then have
\begin{equation}\label{eq:4}
\langle i|H|1\rangle=\sqrt{N}t,
\end{equation}
i.e., an electron hops from site 1 to site $i$ with a matrix element which is 
a factor $\sqrt{N}$ larger than in the one-particle case\cite{Ce}. This is due 
to the fact that any of the $N$ spin up electrons can hop from site 1 to $i$. 
We next construct states $|ij\rangle$ ($j\ne 1$) where a nearest neighbor $j$ 
of $i$ has an extra electron. The corresponding matrix element is $\sqrt{N}t$. 
However, the states $|ij\rangle$ depend on which site $i$ had an extra electron
in the intermediate state. We use the analogy with the one-electron case, 
although it is not rigorous, since in the many-body case there is a string of 
reduced spins along the path of the moving additional occupancy. In the 
one-electron case, the hopping of an electron lowers the energy by 
$\varepsilon_b$, where $\varepsilon_b$ is the bottom of the band. The close 
analogy between the one- and many-particle problems then suggests
\begin{equation}\label{eq:5}
E(NM+1)\approx E(NM)+NU+\sqrt{N}\varepsilon_b,
\end{equation}
with an extra factor $\sqrt{N}$ in front of $\varepsilon_b$. Using a similar 
result for $E(NM-1)$ we obtain
\begin{equation}\label{eq:6}
E_g=U-\sqrt{N}(\varepsilon_t-\varepsilon_b)\equiv U-\sqrt{N}W,
\end{equation}
where $\varepsilon_t$ ($=-\varepsilon_b$ in this model) is the top of the 
one-particle band. This can be generalized to the case where all the hopping 
integrals $t_{im,jm'}\equiv t$ for $i$ and $j$ nearest neighbor sites. For 
$N=1$ we obtain $E_g=U-W$. If this result is extrapolated to smaller  $U$, 
it predicts a  Mott-Hubbard transition for $U/W\sim 1$, close to previous 
results. Eq.\ (\ref{eq:6}) suggests, however, that for a degenerate system the 
transition takes place for a larger ratio $\sim\sqrt{N}$.

To test this, we have performed exact diagonalization calculations for a 
Hubbard model with a six site ($3*2$) lattice with ($N=2$) and without ($N=1$) 
orbital degeneracy. Table \ref{tableI} compares the exact band gap with the  
estimate $U-\sqrt{N}W$. The exact results agree very well with the simple 
estimate for large $U$, confirming the  estimate in the limit it was made. 
For intermediate $U$ the agreement is less good. The reason could be that
the gap is finite for a finite system or that the extrapolation of the 
estimate to intermediate $U$ is invalid. To check this, we consider larger 
systems.

To study A$_3$C$_{60}$ we have considered a model, which includes the 
three-fold degenerate, partly occupied $t_{1u}$ level. We include the on-site 
Coulomb interaction $U$ and the hopping integrals $t_{im,jm'}$ between the 
molecules. This leads to the Hubbard-like model
\begin{eqnarray}\label{eq:7}
H=&&\sum_{i\sigma}\sum_{m=1}^3 \varepsilon_{t_{1u}}n_{i\sigma m}
    +\sum_{<ij>\sigma mm'}                           \nonumber
         t_{ijmm'}\psi^{\dagger}_{i\sigma m} \psi_ {j\sigma m'} \\
+ && U\sum_i\sum_{\sigma m < \sigma'm'}n_{i\sigma m}n_{i\sigma'm'},
\end{eqnarray}
where the sum $<ij>$ is over nearest neighbor sites. We have used a fcc lattice.
The hopping integrals $t_{imjm'}$ have been obtained from a tight-binding
parametrization\cite{Orientation,Satpathy}. The molecules are allowed to 
randomly take one of two orientations in accordance with experiment\cite{%
Stephens}, and the hopping integrals are chosen so that this orientational 
disorder is included\cite{MazinAF}. The band width of the infinite system is 
$W=0.63$ eV. The model neglects multiplet effects. The inclusion of these 
effects may favor antiferromagnetism, which could reduce the critical value 
$U_c/W$ for a Mott-Hubbard transition.

We first consider a cluster of four C$_{60}$ molecules, for which the 
Hamiltonian (\ref{eq:7}) can be diagonalized exactly. The results
are shown in Table \ref{tableI} for the cases of three (A$_{3}$C$_{60}$)
electrons per site. Except for small values of $U$, the gap $E_g$ is actually  
{\sl smaller} than than $U-\sqrt{3}W$, which may be due to the hopping being 
more complicated than discussed above. The importance of the degeneracy is, 
however, clear. 

To study the Mott-Hubbard transition, we consider larger systems. We use a 
diffusion Monte Carlo method, which we have developed along the lines of 
ten Haaf {\it et al.}\cite{ten Haaf}. Starting from a trial function 
$|\Psi_T\rangle$, we calculate 
\begin{equation}\label{eq:9}
|\Psi^{(n)}\rangle = \lbrack 1-\tau(H-w) \rbrack^n |\Psi_T\rangle 
                \equiv F^n |\Psi_T\rangle,
\end{equation}

\noindent
\begin{table}
\caption[]{\label{tableI}
  Band gap as a function of $U$ and $N$ compared with $U-\sqrt{N}W$ for a 
  six-site model ($N=1$ and $N=2$) and a cluster of four C$_{60}$ molecules. 
  W=0.6 for the six atom cluster and 0.58 for the C$_{60}$ cluster. $E_g(n)$ 
  is the band gap for $n$ electrons per site. All energies in eV.}
\begin{tabular}{ccccccc}
\multicolumn{1}{c}{$U$ }&
\multicolumn{2}{c}{$N=1$ (6-site)}&
\multicolumn{2}{c}{$N=2$(6-site)} &                      
\multicolumn{2}{c}{$N=3$(4C$_{60}$)}\\                  
 &$E_g(1)$ & $U-W$  &  $E_g(2)$   & $U-\sqrt{2}W$ & 
  $E_g(3)$    &       $U-\sqrt{3}W$  \\
\tableline
10  & 9.35  & 9.40   & 9.06    & 9.15 & 8.39     &   9.00   \\
 5  & 4.39  & 4.40   & 4.12    & 4.15 & 3.48     &   4.00   \\
 3  & 2.44  & 2.40   & 2.18    & 2.15 & 1.64     &   2.00   \\
 2  & 1.48  & 1.40   & 1.24    & 1.15 & 0.90     &   1.00   \\
 1  & 0.54  & 0.40   & 0.35    & 0.15 & 0.46     &   0.00   \\
\end{tabular}
\end{table}
\begin{table}
\caption{
  The total energy for four C$_{60}$ molecules with $M$ electrons and $U=1$
  according to the variational (VMC) and diffusion (DMC) Monte Carlo 
  calculations compared with the exact results. The Gutzwiller parameter is 
  given by $g$. The statistical error is estimated to be two units in the 
  last digit.}
\begin{tabular}{ccccc}
   M    &  g    &   VMC      &    DMC     &   Exact   \\
\tableline
   11   & 0.45  & \ 8.4970   & \  8.4677  &                \\
   11   & 0.50  & \ 8.4842   & \  8.4677  & \ 8.4649       \\
   11   & 0.55  & \ 8.4873   & \  8.4684  &                \\
   12   & 0.45  &  10.7364   &   10.7014  &                \\
   12   & 0.50  &  10.7226   &   10.7009  &  10.6994       \\
   12   & 0.55  &  10.7249   &   10.7012  &                \\
   13   & 0.45  &  13.4363   &   13.3990  &                \\
   13   & 0.50  &  13.4211   &   13.3988  &  13.3973       \\
   13   & 0.55  &  13.4233   &   13.3991  &                \\
  $E_g$ & 0.45  &\  0.4605   & \  0.4639  &                \\
  $E_g$ & 0.50  &\  0.4601   & \  0.4647  & \ 0.4634       \\
  $E_g$ & 0.51  &\  0.4608   & \  0.4651  &                \\
\end{tabular}
\label{tableII}
\end{table}

\noindent
where $w$ is an estimate of the ground-state energy. If $\tau$ is sufficiently 
small and $|\Psi_T\rangle$ is not orthogonal to the ground-state,
$|\Psi^{(n)}\rangle$ converges to the ground-state as $n\to\infty$. 
$|\Psi^{(n)}\rangle$ is calculated in a Monte Carlo approach. Although this 
approach in principle is exact, it suffers from the usual sign problem.
We therefore use the lattice equivalent of a fixed node approximation\cite{%
ten Haaf}, and introduce an effective Hamiltonian for which the 
nondiagonal terms are zero if
\begin{equation}\label{eq:10}
\langle\Psi_T|R\rangle\,\langle R|F|R'\rangle\,\langle R'|\Psi_T\rangle \;<0,
\end{equation}
where $R$ represents the coordinates of all the electrons. A term is added to 
the diagonal part of the effective Hamiltonian, so that its energy is an upper 
bound to the exact energy, with the  upper bound agreeing with the exact energy
if $|\Psi_T\rangle$ is the exact wave function\cite{ten Haaf}.

For the trial function we make the Gutzwiller Ansatz
\begin{equation}\label{eq:8}
  |\Psi_T\rangle=g^{n_d}|\Psi_0\rangle                            
\end{equation}
where $g$ is a Gutzwiller parameter and the power $n_d$ is the number of 
double occupations $n_d=\sum_i\sum_{\sigma m < \sigma'm'}n_{i\sigma m}
n_{i\sigma'm'}$ in the Slater determinant $|\Psi_0\rangle$. To construct 
$|\Psi_0\rangle$ we solve the Hamiltonian (\ref{eq:7}) in the Hartree-Fock 
approximation, replacing $U$ by a variational parameter $U_0$. Depending on 
$U_0$ the trial function is paramagnetic or antiferromagnetic.           

As a test, we study a cluster of four C$_{60}$ molecules. We calculate the 
expectation value of $H$ for $|\Psi_T\rangle$ in a variational Monte Carlo 
(VMC) approach and perform a diffusion calculation (DMC). The results are 
compared with the exact results in Table \ref{tableII}. The agreement is very 
good, with an error of 0.003 eV or less in the DMC calculation for the 
parameters considered.  

To obtain an efficient extrapolation to infinite systems, we correct for 
finite size effects\cite{cep87}. For the present problem, there should be a 
gap for a finite system even for a small $U$. We can see this by distributing 
the charge uniformly over all the levels and calculate the electrostatic energy.
This gives a contribution $U/M$ to the gap\cite{gap}, where $M$ is the number 
of molecules. We then introduce 
\begin{equation}\label{eq:11}
\tilde E_g\equiv E_g-{U\over M}-E_g(U=0),              
\end{equation}
where $E_g(U=0)$ is the gap in the one-particle spectrum. We may then expect 
$\tilde E_g$ to be rather independent of $M$ for itinerant systems. Both $U/M$ 
and $E_g(U=0)$ go to zero for $M \to \infty$, so the subtracted quantities 
improve the convergence but do not change the $M\to \infty$ result.       

\begin{figure}
\centerline{\epsfxsize=3.1in \epsffile{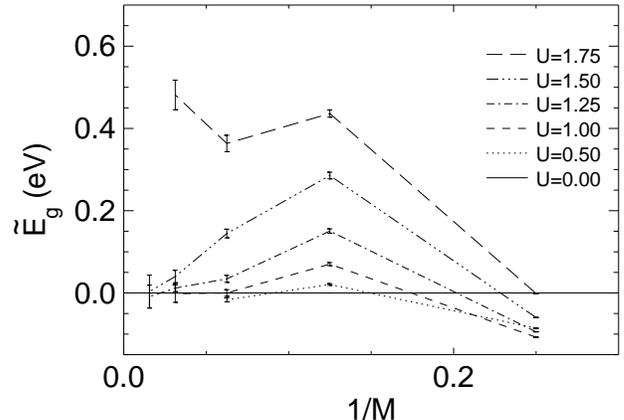}}
\noindent
\caption[]{\label{fig1} 
   $\tilde E_g$ (Eq.\ (\ref{eq:11}) for different $U$ and  as a function of
   $1/M$, where $M$ is the number of molecules. The band width varies between 
   0.58 eV ($M=4$) and 0.63 eV ($M=\infty$).}
\end{figure}

In Fig.~\ref{fig1} we show results for $\tilde E_g$ that suggest a Mott-Hubbard
transition for $U$ between 1.5 and 1.75 eV. This corresponds to a critical 
ratio $U_c/W\sim$ 2.5. This is substantially smaller than the ratio 4 found 
by Lu\cite{Lu}. For a fully frustrated system with infinite dimension but 
without orbital degeneracy $U_c/W=1.5$ was obtained\cite{Mott-Hubbard}.
Multiplying by the degeneracy factor $\sqrt{3}$ leads to the ratio 
$\sqrt{3}\times 1.5=2.6$ in good agreement with our results. It is interesting 
to note that for U=1.75, which gives an insulator, an antiferromagnetic trial 
function results in the lowest energy for the larger systems.                 

The inclusion of higher subbands may increase $U_c/W$ further. In the large 
$U$, half-filled case, the hopping to a higher subband costs an extra 
``diagonal'' energy $U+\Delta \varepsilon$, where $\Delta \varepsilon$ is the 
energy separation of the bands. The corresponding energy cost in the states 
with one extra electron or hole is only $\Delta \varepsilon$. The hopping to 
higher subbands should therefore lower the energy more in the latter two 
states, which decreases the gap. Simultaneously, the one-particle width of the 
$t_{1u}$ band due to hopping into higher levels on neighboring molecules is 
{\it reduced } by a few per cent\cite{Satpathy}. Only bands with 
$\Delta\varepsilon\lesssim U$ contribute to the reduction of the band gap. 
A simple model suggests that this could increase
$U_c/W$ by 0.1-0.2 for A$_3$C$_{60}$. For $3d$, $4f$ or $5f$ compounds 
($N=5$ or 7), a crystal-field splitting may lower the degeneracy. The higher 
crystal-field split states cannot, however, be neglected when determining 
$U_c/W$, unless the crystal-field splitting is much larger than the Coulomb 
interaction.                     

The average Coulomb $U$ has been estimated to be 1.6 eV\cite{Sawatzky} and 
$1.4\pm0.2$ eV\cite{Bruhwiler} from Auger experiments for the occupied 
orbitals. A slightly smaller (0.2 eV) value was found for the highest occupied 
orbital ($h_u$)\cite{Sawatzky}, and $U$ may be somewhat further reduced for 
the higher-lying  $t_{1u}$ orbital. The reduced screening in the surface layer 
is estimated to increase $U$ in this layer by 0.3 eV\cite{Antropov}. The 
surface sensitivity of Auger may therefore slightly overestimate $U$.
Theoretically, $U$ has been estimated to be in the range 0.8-1.3 eV\cite{%
UAntropov}. Band structure calculations have found the band width 
0.41 eV\cite{Martins}, 0.6 eV\cite{Pickett} and 0.65 eV\cite{WAntropov} 
for K$_3$C$_{60}$. Many-body effects were found to increase the band width 
by about 30$\%$ in the GW approximation for undoped C$_{60}$\cite{Louie}. 
The many-body effects treated in the Hubbard model, due to the interactions 
in the $t_{1u}$ band, are not included in the GW calculation, since the 
$t_{1u}$ band was empty in the GW calculation. Using the GW enhancement of 
the band width is therefore not a double-counting. For Rb$_3$C$_{60}$ the 
band width may be reduced by about 15$\%$\cite{Satpathy}, compared with 
K$_3$C$_{60}$. This leads to a likely ratio $U/W$ in the range 1.5-2.5. 
In view of our QMC results, this is probably too small to cause a Mott-Hubbard 
transition for A$_3$C$_{60}$.                                     

One could ask what are the effects of $U>W$ in a metal. The system can remain 
a metal in spite of the large $U/W$ thanks to the hopping via several channels
(Eq.\ (\ref{eq:3})). As in the impurity problem\cite{Ce}, this may favor the 
formation of a singlet state. In A$_3$C$_{60}$, which has an odd number of 
electrons per site, the singlet is consistent with a correlated non-magnetic 
state, which must be metallic if it obeys the Luttinger theorem\cite{lut}. 
It is suggestive that such singlet correlations could tend to promote 
superconductivity in these systems.

While A$_3$C$_{60}$ are  metals, A$_4$C$_{60}$ are insulators\cite{A4C60}, 
although band structure calculations predict that they should be metals%
\cite{Erwin}. A$_4$C$_{60}$ is therefore not a band insulator, and this  
raises interesting questions about the difference between A$_3$C$_{60}$ and 
A$_4$C$_{60}$. The presence of four electrons per molecule in A$_4$C$_{60}$
favors the formation of a static Jahn-Teller effect, which may open up a 
gap\cite{Jahn-Teller}, in particular if electron correlation effects are also 
included. The electron-phonon coupling should reduce the effective hopping
for both A$_3$C$_{60}$ and A$_4$C$_{60}$. One may also speculate that the gap 
in A$_4$C$_{60}$ is a consequence of the interactions in a case with an even 
number of electrons per site. This could be a correlation induced gap with no 
development of magnetic moments, i.e., a different type of order than in other 
correlated Mott insulators such as NiO. Besides A$_4$C$_{60}$ is bipartite 
while A$_3$C$_{60}$ is not, which may favor the development of ordered phases 
(anti-ferromagnetic or other types of order) and the opening of a gap in 
A$_4$C$_{60}$. Finally, the Fermi surface of A$_4$C$_{60}$ has almost perfect 
nesting, which may favor a charge density wave and the opening of a 
gap\cite{Erwin}.
 
To summarize, we have shown that the orbital degeneracy increases the value of 
$U/W$ where the Mott-Hubbard transition takes place, due to the more efficient 
hopping. This puts the critical ratio for A$_3$C$_{60}$ at the upper range of
ratios $U/W$ estimated for doped C$_{60}$ compounds and makes it likely that 
A$_3$C$_{60}$ (A=K, Rb) is on the metallic side. Although experimental samples 
of A$_3$C$_{60}$ may be nonstoichiometric, it is therefore not necessary to 
assume so to explain the metalicity. The large value of $U/W$ may nevertheless 
lead to substantial correlation effects, which may have a large influence on 
certain properties. Most systems of interest have orbital degeneracy\cite{%
Bickers}. Therefore our results should be relevant also for many other systems 
close to a Mott-Hubbard transition.

We would like to thank  D.F.B.~ten Haaf, H.J.M.~van Bemmel, J.M.J.~van Leeuwen,
and W. van Saarloos for helpful conversations in the early stages of this work.
A.~Burkhardt and A.~Schuhmacher provided invaluable help in adapting the 
programs for parallel computation. This work was supported in part by the the 
NSF under grants DMR-94-22496 and DMR-89-20538.

\end{multicols}

\begin{thebibliography}{*}
\bibitem{gap}
Performing a Hartree-Fock like calculation for this 
uniform system, would give a contribution $5U/(6M)$.
\end{thebibliography}
\end{document}